# An Exact Local Hybrid Monte Carlo Algorithm for Gauge Theories*

A. D. Kennedy and K. M. Bitar

Supercomputer Computations Research Institute
Florida State University
Tallahassee, FL 32306–4052, USA
Internet: {adk,kmb}@scri.fsu.edu

**Abstract**

We introduce a new Monte Carlo method for pure gauge theories. It is not intended for use with dynamical fermions. It belongs to the class of Local Hybrid Monte Carlo (LHMC) algorithms, which make use of the locality of the action by updating individual sites or links by following a classical mechanics trajectory in fictitious time. We choose to update a one-parameter subgroup of the gauge field on each link of the lattice, and the classical trajectory can be found in closed form in terms of elliptic functions for this case. We show that this gives an overrelaxation algorithm with a tunable parameter which, unlike some previous methods, does not require the numerical integration of the equations of motion.

---

# 1 LHMC FOR FREE FIELD THEORY

Let us first quickly review the Local Hybrid Monte Carlo (LHMC) and Adler's over-relaxation (AOR) algorithms [1] for free field theory. See [2] for further references. Consider the Gaussian model defined by the free field action

$$S(\phi) = \frac{1}{2} \sum_x \left\{ \sum_\mu \left( \partial_\mu \phi(x) \right)^2 + m^2 \phi(x)^2 \right\}.$$

A single-site AOR update with parameter $\zeta$ replaces $\phi(x)$ by

$$\phi'(x) = (1 - \zeta)\phi(x) + \frac{\zeta \mathcal{F}}{\omega^2} + \frac{\sqrt{\zeta(2 - \zeta)}}{\omega} \eta,$$

where $\omega^2 \equiv 2D + m^2$, $\mathcal{F} \equiv \sum_{|x-y|=1} \phi(y)$, and $\eta$ is a Gaussian random number.

The lattice may be updated using a checkerboard scheme.

- For $\zeta = 1$ this is the heatbath (HB) algorithm, because $\phi'(x)$ does not depend upon $\phi(x)$.

  The exponential autocorrelation time is

  $$\tau_{\text{HB}} = \frac{D}{m^2 + p^2} \qquad (p^2 \to 0),$$

  which corresponds to $z = 2$.

- If we adjust $\zeta$ so as to minimize the autocorrelation time we find that

  $$\zeta = 2 - \frac{2m}{\sqrt{D}} + O(m^2), \quad \tau_{\text{AOR}} \approx \frac{\sqrt{D}}{2m};$$

  which gives $z = 1$.

Consider the LHMC algorithm applied to the Gaussian model: We introduce the Hamiltonian $H = \frac{1}{2}p^2 + S(\phi)$ on "fictitious" phase space. The corresponding equation of motion are

$$\ddot{\phi}_x = -\omega^2 \phi_x + \mathcal{F},$$

whose solution in terms of the Gaussian distributed random initial momentum $p_x$ and the initial field value $\phi_x$ is

$$\phi_x(t) = \phi_x \cos \omega t + \frac{1 - \cos \omega t}{\omega^2} \mathcal{F} + \frac{p_x}{\omega} \sin \omega t,$$

which is exactly the AOR update considered before if we identify $\zeta \equiv 1 - \cos \omega t$ and $p_x \equiv \eta$.

If we use this exact solution of the equations of motion to generate candidate configurations for LHMC then the acceptance rate will be unity.

# 2 LHMC FOR PURE GAUGE THEORY

We now turn to the case of pure gauge theory. The Wilson action is

$$S(U_\ell) = -\beta \operatorname{Re} \frac{\operatorname{Tr} V}{\operatorname{Tr} 1}$$

where the matrix $V \equiv \sum_{\mathcal{P} \in \partial^* \ell} U_\mathcal{P}$.

Our update will be $U_\ell \to g U_\ell$ where $g$ is a group element.

There is no reason why all the degrees of freedom on a link need to be updated simultaneously; we are free to update just a single degree of freedom at a time as long as we can guarantee ergodicity. Therefore we shall restrict $g$ to lie in a one-dimensional subgroup.

For $U$ in the fundamental representation of $SU(N)$ or $SO(N)$ any group element can be written as a conjugate of

$$\begin{bmatrix} e^{i\theta} & & \\ & e^{-i\theta} & \\ & & \ddots \end{bmatrix} \quad \text{or} \quad \begin{bmatrix} \cos\theta & \sin\theta & \\ -\sin\theta & \cos\theta & \\ & & \ddots \end{bmatrix}$$

respectively, so updating with just this subgroup will be ergodic if random gauge transformations are carried out from time to time. In other words, selecting an element from these subgroups is equivalent to choosing a completely general group element up to gauge transformations.

If we introduce the parameters

$$\alpha e^{i\delta} \equiv \begin{cases} V_{11} + V_{22}^* & \text{for } SU(N) \\ (V_{11} + V_{22}) + \\ \quad i(V_{12} - V_{21}) & \text{for } SO(N) \end{cases}$$

then we may write the action in the simple form

$$S(\theta) = \frac{\beta \alpha}{N} \{1 - \cos(\theta + \delta)\}.$$

As usual, we introduce a fictitious momentum $p$ and a Hamiltonian

$$H(p, \theta) = \frac{1}{2} p^2 + S(\theta) = \frac{1}{2} p^2 + V_0 \sin^2\left(\frac{\theta + \delta}{2}\right)$$

where $V_0 \equiv 2\beta\alpha/N$ is the maximum potential energy of the "pendulum".

Our LHMC algorithm consists of repeatedly applying the following steps in any order:

- Refresh the momentum $p$ from a Gaussian heatbath;

- Integrate the equations of motion for a time $\tau(E)$ and then reverse the sign of the final momentum;
- Perform a random gauge transformation.

The only subtlety is that we can allow the trajectory length $\tau$ depend upon the constants of motion (specifically the energy $E$) because for an exact solution of the equations of motion this still satisfies detailed balance.

We now have to solve the equations of motion, which are

$$\ddot{\theta} = -V_0 \sin\left(\frac{\theta + \delta}{2}\right) \cos\left(\frac{\theta + \delta}{2}\right)$$

subject to the initial conditions $\theta(0) = 0$ and $p(0) = p_0$.

It is convenient to introduce a new variable $y \equiv \sin\frac{\theta+\delta}{2}$ and a rescaled time $u \equiv t\sqrt{E/2}$, whence

$$\frac{dy}{du} = \sqrt{(1-y^2)(1-k^2y^2)}$$

where the *parameter* $k \equiv \sqrt{V_0/E}$. The solution of this equation is the Jacobian elliptic function

$$y = \mathrm{sn}(u\,\mathrm{sgn}\,p_0 + u_0, k)$$

where $\mathrm{sn}\,u_0 = \sin\frac{\delta}{2}$.

Remember that we can choose any value $\tilde{u}(k)$ for the trajectory length. One particularly interesting choice is to take the trajectory length to be a fixed fraction of half the period of an orbit, $\bar{u}$, which corresponds to overrelaxation;

$$\tilde{u}(k) = \left(1 - \frac{1}{\xi}\right)\bar{u},$$

where $\xi$ is (proportional to) the longest relevant correlation length of the gauge theory.

A pendulum exhibits two different types of motion depending on the value of the parameter $k$ (for $k = 1$ the solution is a Sine–Gordon instanton)

$$\bar{u} = \begin{cases} 2K(k) & \text{if } k < 1,\ E > V_0 \\ \frac{2}{k}K\left(\frac{1}{k}\right) & \text{if } k > 1,\ E < V_0, \end{cases}$$

where $K$ is the *complete elliptic integral*

$$K(k) \equiv \int_0^1 \frac{dy}{\sqrt{(1-y^2)(1-k^2y^2)}}.$$

With this choice of trajectory length the value of $\theta$ selected may be written as

$$\tilde{\theta} = 2\,\mathrm{am}\left(u_0 - \frac{\bar{u}\,\mathrm{sgn}\,p_0}{\xi}, k\right),$$

where the *amplitude* $\mathrm{am}\,u \equiv \sin^{-1}\mathrm{sn}\,u$.

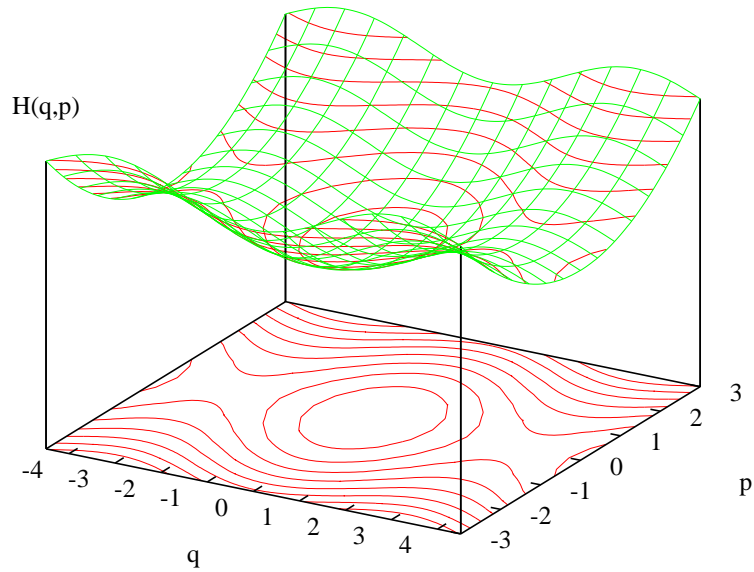

Figure 1: The Hamiltonian as a function on phase space (with $q \equiv \theta$) with the corresponding classical trajectories.

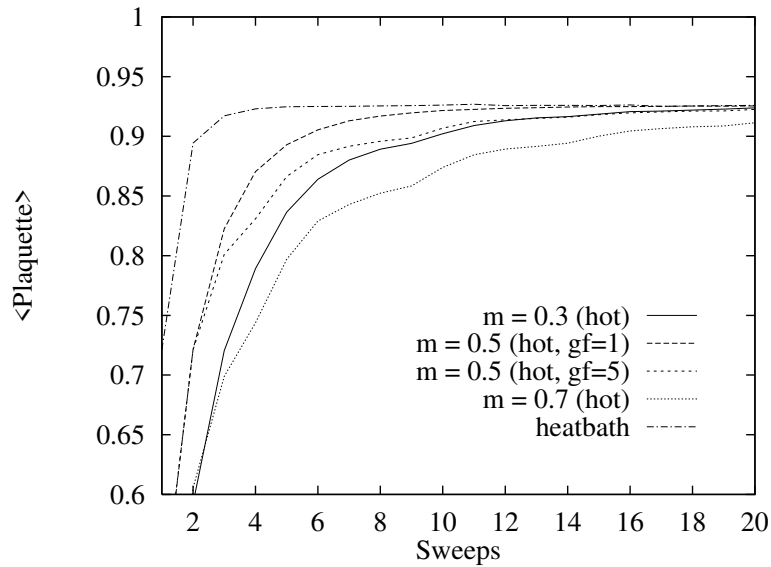

Figure 2: *Preliminary* results for two-dimensional $SU(2)$ on an $128^2$ lattice at $\beta = 20.0$. Convergence for various values of $m = 1/\xi$ and for hot and cold starting configurations is demonstrated. Studies of the dynamical critical behaviour are in progress.

# 3 CONCLUSIONS

We may hope that the LHMC algorithm, with optimally chosen trajectory lengths, will exhibit a dynamical critical exponent $z \approx 1$, if the results from free field theory prove to be applicable to pure gauge theories. In this regard it is similar to the Hybrid Overrelaxation algorithm analysed by Wolff.

The advantages of the LHMC method as compared to previous overrelaxation methods for pure gauge theories is that it allows a true tunable overrelaxation parameter without the need for explicit numerical integration.

# ACKNOWLEDGEMENTS

This research was supported by the Florida State University Supercomputer Computations Research Institute which is partially funded by the U.S. Department of Energy through contracts #DE-FC05-85ER250000 and #DE-FG05-92ER40742.